\documentstyle[aps,twocolumn,epsf]{revtex}
\begin{document}
\title{Multiple forms of intermittency in PDE dynamo models}
\author{
Eurico Covas\thanks{E-mail: eoc@maths.qmw.ac.uk}\thanks{Web: http://www.maths.qmw.ac.uk/$\sim$eoc} and
Reza Tavakol\thanks{E-mail: reza@maths.qmw.ac.uk}}
\address{
Astronomy Unit,
Mathematical Sciences,
Queen Mary and Westfield College,
Mile End Road,
London E1 4NS,
United Kingdom.
}
\maketitle
\begin{abstract}
We find concrete evidence for the presence of crisis--induced and
Pomeau--Manneville Type-I intermittencies in an axisymmetric PDE mean--field
dynamo model. These findings are
of potential importance for two different reasons. Firstly, as far as
we are aware, this is the first time detailed evidence has been
produced for the occurrence of these types of intermittency for such
deterministic PDE models. And secondly, despite the rather idealised
nature of these models, the concrete evidence for the
occurrence of more than one type of intermittency in such models
makes it in principle possible that different types of intermittency
may occur in different solar-type stars or even in the same star over
different epochs. In this way a {\it multiple intermittency framework} may turn
out to be of importance in understanding the mechanisms responsible for
grand-minima type behaviour in the Sun and solar-type stars and in
particular in the interpretation of the corresponding observational and
proxy evidence.
\end{abstract}
\section{Introduction}
Intermittency has been observed in a variety of real settings as well in a
vast number of numerical models. A great deal of effort
has therefore gone into understanding these modes of behaviour in the context
of deterministic dynamical systems theory. These studies have demonstrated the
existence of a number of different types of intermittency (such as
Pomeau--Manneville \cite{pomeau}, Crisis \cite{grebogietal},
On-off \cite{ashwinplattspiegel} intermittencies), each with their own
associated signatures and scalings.  Many of these forms of intermittency have
in turn been concretely shown to be present in experiments and numerical
studies of dynamical systems in a variety of settings (see
\cite{ottbook93,numerics,1overf}
and references therein).

An important potential domain of applicability of such behaviour arises in
understanding the mechanisms underlying the intermediate time scale
variability in the Sun \cite{eddy76-weiss} - the occurrence of the so called {\it Maunder} or
{\it grand minima} - during which solar activity (as deduced from the sunspot
numbers) was greatly diminished \cite{eddy76-weiss,beer98}. This behaviour is
also confirmed by evidence coming from the analysis of proxy data
\cite{stuiver}.
There is also some evidence for similar types of variability in solar-type
stars \cite{weiss94etc}.

The idea that some type of dynamical intermittency may under pin the
grand minima type variability in the sunspot record (the {\it
intermittency hypothesis} \cite{tavakol-covas99}) goes back at least to
the late 1970's \cite{tavakol78,zeldovichetal83,spiegel85etc}.  This
idea has been the subject of intense study over the recent years and
has involved the employment of various classes of dynamo models,
including ordinary differential equations (ODE) (e.g.\
\cite{zeldovichetal83,jonesetal85} as well as partial differential
equations (PDE) models (e.g.\ \cite{schmitt,tt,tobias96-97,inout}).  In
addition to the phenomenological evidence for the presence of
intermittent-type behaviours in dynamo models
\cite{schmitt,tt,tobias96-97,brooke98,brooke97}, concrete evidence has
recently been found for the presence of particular types of
intermittency in both ODE dynamo models \cite{covas97c,covas97d} as
well as a recently discovered generalisation of on-off intermittency,
referred to as {\it in-out} intermittency \cite{act99}, in PDE models
\cite{inout}.

Here we wish to report concrete evidence for the occurrence of two other types
of intermittency, namely the crisis--induced and Pomeau--Manneville Type-I
intermittencies, in PDE mean--field dynamo models.
The organisation of the paper is as follows. In Sec.\ II
we briefly introduce the model studied here. Sec.\ III summarises our
evidence demonstrating the presence of these types of intermittencies
in this model and finally in Sec.\ IV we draw our conclusions.
\section{Model}
Ideally one would wish to employ the full 3-dimensional dynamo models
with a minimum number of approximations and simplifying assumptions.
Despite a number of important recent attempts
\cite{gilman83,nordlundetal92,brandenburgetal96},
the difficulty of dealing with small scale turbulence
makes a detailed and extensive self consistent study of such fully
turbulent regimes in stars still computationally impractical
(see e.g.\ \cite{nordlundetal92,cattaneoetal91,mossetal95,vainshtein}.

In view of this an alternative approach in studies of
stellar dynamos has been to employ mean--field models
\cite{jonesetal85,schmitt,tobias96-97,brooke98,steenbecketc,brandenburg89}.
We should mention that there is an ongoing debate regarding
the nature and realistic value of such models
\cite{vainshtein}. Nevertheless, 3-D
turbulence simulations do seem to produce magnetic fields whose global
properties (such as field parity and time dependence)
are similar to those expected
from corresponding mean--field dynamo models \cite{brandenburg99b}.
In this way mean--field dynamo models seem to reproduce
certain features of the more complicated models and allow the study of certain
global properties of magnetic fields in the Sun
and solar-type stars (see for example \cite{brandenburg99b,brandenburg99a}).

The standard mean--field dynamo equation is given by
\begin{equation} \label{dynamo}
\frac{\partial {\bf B}}{\partial t}=
\nabla \times \left( {\bf u} \times {\bf B} + \alpha {\bf B} - \eta_t
\nabla \times {\bf B} \right),
\end{equation}
where ${\bf B}$ and ${\bf u}$ are the mean magnetic field and mean velocity
respectively and the turbulent magnetic diffusivity $\eta_t$ and the
coefficient $\alpha$ arise from the correlation of small scale turbulent
velocities and magnetic fields ($\alpha$ effect) \cite{krause}.
We consider the usual algebraic form of $\alpha$--quenching namely
\begin{equation}\label{alpha_a}
\alpha=\frac{\alpha_0 \cos \theta}{1+ |{\bf B}|^2},
\end{equation}
where $\alpha_0={\rm constant}$ and $\theta$ is the co-latitude.

We solve Eq.\ \ref{dynamo} in an axisymmetric configuration and in
the following, as is customary \cite{brandenburg89}, we shall discuss the
behaviour of the solutions by monitoring the total magnetic
energy, $E={1\over2\mu_0}\int{\bf B}^2dV$, where $\mu_0$ the
induction constant, and the integral is taken over the 
dynamo region. We split $E$ into two parts,
$E=E_A+E_S$, where $E_A$ and $E_S$ are
respectively the energies of the antisymmetric and symmetric
parts of the field with respect
to the equator.  The overall parity $P$ is given by
$P=[E_S-E_A]/E$, so $P=-1$
denotes an antisymmetric (dipole-like) pure parity solution
and $P=+1$ a symmetric (quadrupole-like) pure parity
solution.

For the numerical results reported in the following section,
we used a modified version of the axisymmetric dynamo code
of Brandenburg {\em et al.} (1989) \cite{brandenburg89}
employed recently in \cite{cutpaper}.  These models are
constructed from a complete sphere of radius $R$ by removing
an inner concentric sphere of radius $r_0$ and a conical
section of semi-angle $\theta_0$ about the rotation axis,
from both the north and south polar regions (see \cite{cutpaper}
for details of the model and the relevant parameters).  To
test the robustness of the code we verified that no
qualitative changes were produced by employing a finer grid
and different temporal step length (we used a grid size of
$41\,\times\,81$ mesh points and a step length of
$10^{-4}R^2/\eta_t$ in the results presented in this paper).
For the following results we use $C_\Omega=-10^{4}$, which
give the magnitude of the differential rotation and
$\theta_0=45^{\circ}$. The magnitude of the $\alpha$-effect
is given by the dynamo parameter $C_{\alpha}$.

In the next section we show in turn
concrete evidence for the occurrence of
crisis--induced and Pomeau--Manneville Type-I
intermittencies.
\section{Results}
\subsection{Crisis--induced Intermittency}
As far as their detailed underlying mechanisms and temporal signatures
are concerned, crises come in three varieties \cite{grebogietal}.
Here we shall be concerned with only one of these types, referred to as ``attractor merging
crisis'', whereby as a system parameter is varied, two or more chaotic
attractors merge to form a single attractor.
There are both
experimental and numerical evidence for this type of intermittency (see for
example \cite{grebogietal,ottbook93} and references
therein). In particular, this type of behaviour has been discovered in a 6-dimensional
truncation of mean--field dynamo models \cite{covas97c}.

\begin{figure}[!htb]
\centerline{\def\epsfsize#1#2{0.46#1}
\epsffile{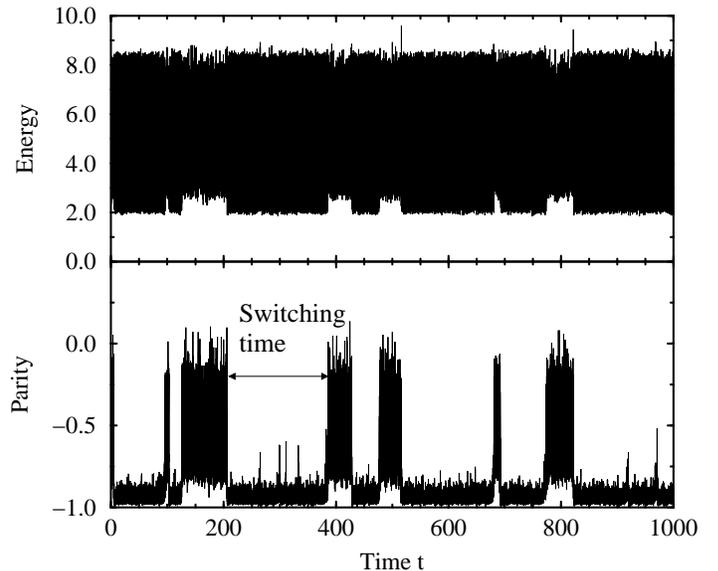}}
\caption{\label{Crisis.series}
Example of crisis induced intermittency in a shell dynamo with a cut,
with $r_0=0.2$, $C_\alpha=25.202$, $C_\Omega=-10^4$ and $\theta_0=45^{\circ}$.
}
\end{figure}

Fig.\ \ref{Crisis.series} shows the plots of the energy and parity for the
above model as a function of time, calculated with $r_0=0.2$ and $C_\alpha = 25.202$
which show a bimodal behaviour, switching
intermittently between two different chaotic states.

\begin{figure}[!htb]
\centerline{\def\epsfsize#1#2{0.43#1}\epsffile{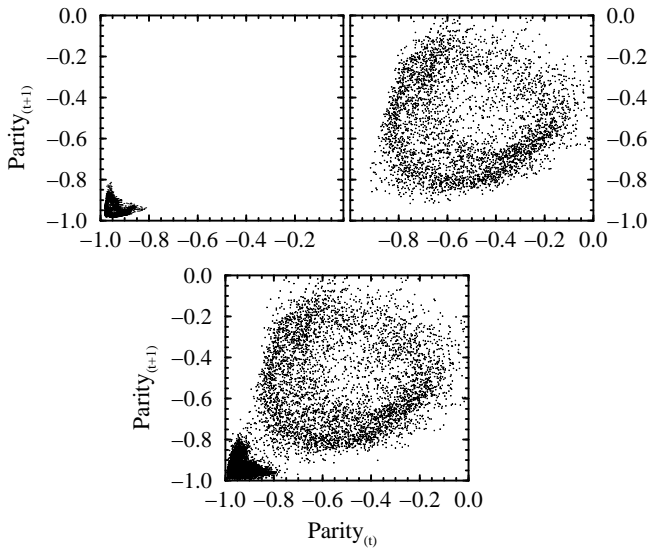}}
\caption{\label{return_map.eps}
Return maps showing the attractors in the PDE model (1) before (top panels) and after the
merging (bottom panel).
Note that as expected the merged attractor
is larger than the superposition of the two previous attractors.}
\end{figure}

To determine the nature of this behaviour more
precisely, we have plotted in
Fig.\ \ref{return_map.eps} the return maps
for the PDE models (1),
showing the attractors before and after
the merging.
As can be seen the resulting merged attractor is, as expected,
larger than the superposition of the
two pre-existing attractors.

These results can be taken as indications
for the presence of crisis--induced intermittency
in this model. To substantiate this further,
we recall that another important signature of this type of intermittency
is the way $\tau$, the average time between switches, scales with
the system parameter, in this case, $C_\alpha$. According to
Grebogi {\em et al.} \cite{grebogietal}, for a large class
of dynamical systems this relation takes the form
\begin{equation} \label{scaling}
\tau \sim \left|C_\alpha-C_\alpha^* \right |^{-\gamma},
\end{equation}
where the real constant $\gamma$ is the critical exponent characteristic
of the system under consideration and $C_\alpha^*$ is the critical value of $C_\alpha$
at which the two chaotic attractors merge.

\begin{figure}[!htb]
\centerline{\def\epsfsize#1#2{0.46#1}
\epsffile{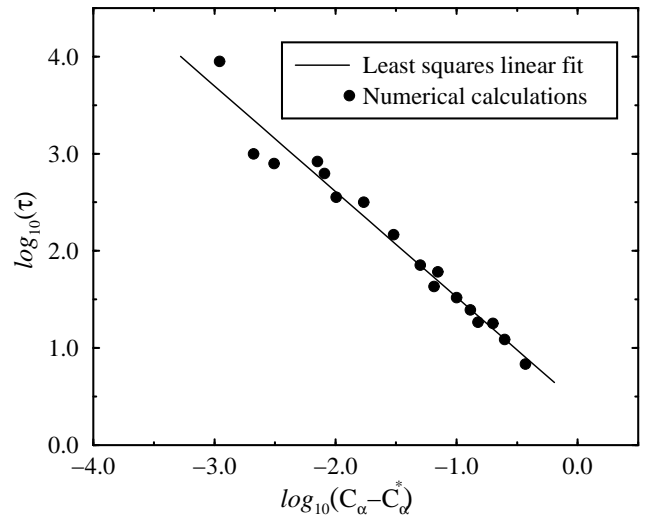}}
\caption{\label{Crisis.statistics}
Scaling of the average times between switches
$\tau$ as a function of $(C_\alpha-C_\alpha^*)$
 for crisis
induced intermittency for the model (\ref{dynamo}).
The slope is found to be
$\gamma=1.08 \pm 0.05$}
\end{figure}

The model under study here is a PDE system which is formally infinite
dimensional. Such PDE models are numerically costly to integrate over
long enough intervals of time (sometimes in excess of 5000 time units)
necessary in order to obtain the scaling of the type (\ref{scaling}).
Furthermore, the demonstration of such scaling requires a precise
determination of the critical value $C_\alpha^*$ which is difficult
since as one approaches this value $\tau$ diverges and the integration
time becomes prohibitive.  Despite these difficulties, we have
succeeded to obtain strong evidence for the presence of such a scaling
as depicted in Fig.\ \ref{Crisis.statistics}, with the corresponding
$\gamma=1.08 \pm 0.05$. Grebogi {\em et al.} \cite {grebogietal}
conjecture that there may be a general tendency for $\gamma$ to be
larger for higher--dimensional attractors. We do have a value of
$\gamma$ higher than the previous one found for a related six
dimensional ODE dynamo model \cite{covas97c} but much lower than the
value range suggested by Grebogi {\em et al.} Therefore, the
conjectured range may need modification for large high--dimensional
systems.

There is also evidence for an enlargement of the final attractor
after merging, as shown by the larger amplitudes of
variation in the parity, in the sense that the parity gets
closer to $-1$ after the merging, as depicted in Fig.\
\ref{return_map.eps}. This helped us
to numerically arrive at a better estimate for the critical value
$C_\alpha^*$.

These indicators, taken together, amount to strong evidence for the
presence of crisis--induced intermittency for this model.
\subsection{Pomeau--Manneville Type-I Intermittency}
This type of intermittency, which is
brought about through
a tangent bifurcation, results in the system switching back and
forth between a ``ghost'' periodic orbit and sudden
bursts of chaotic behaviour \cite{pomeau}.
There are both
experimental and numerical evidence for this type of intermittency
(see for example \cite{numerics,hirsch} and references
therein). In particular this type of behaviour has been discovered in a 12-dimensional
truncation of mean--field dynamo model \cite{covas97d}.

\begin{figure}[!htb]
\centerline{\def\epsfsize#1#2{0.46#1}
\epsffile{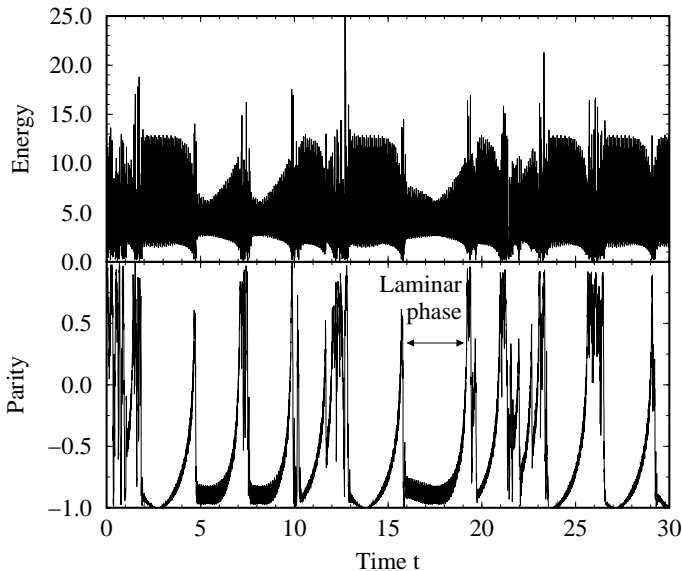}}
\caption{\label{TypeI.series}
Example of Type-I intermittency in a shell dynamo with a cut,
with $r_0=0.7$, $C_\alpha=28.0$, $C_\Omega=-10^4$ and $\theta_0=45^{\circ}$.}
\end{figure}

To demonstrate the presence of this type of intermittency in the above PDE
dynamo model, we have plotted in Fig.\ \ref{TypeI.series} the energy and parity
as a function of time for the parameter values $r_0=0.7$ and $C_\alpha = 28.0 $, which clearly
demonstrates switches between nearly periodic behaviour and sudden bursts. We
note that interestingly the energy in this case shows strong modulation which
could be of interest in accounting for the occurrence of grand type minima
in sunspot activity.

Another signature of this type of intermittency is provided by the specific
characteristics of its corresponding power spectrum.  By employing finite
dimensional maps\,\cite{1overf}, it has been shown that the corresponding
spectra have a broad-band feature whose shape obeys approximately the
inverse-power law $1/f$ for $f>f_s$, where $f_s$ is the saturation frequency.
Below this frequency there is a flat plateau induced by noise that causes
arbitrarily long laminar phases to become finite.

\begin{figure}[!htb]
\centerline{\def\epsfsize#1#2{0.46#1}
\epsffile{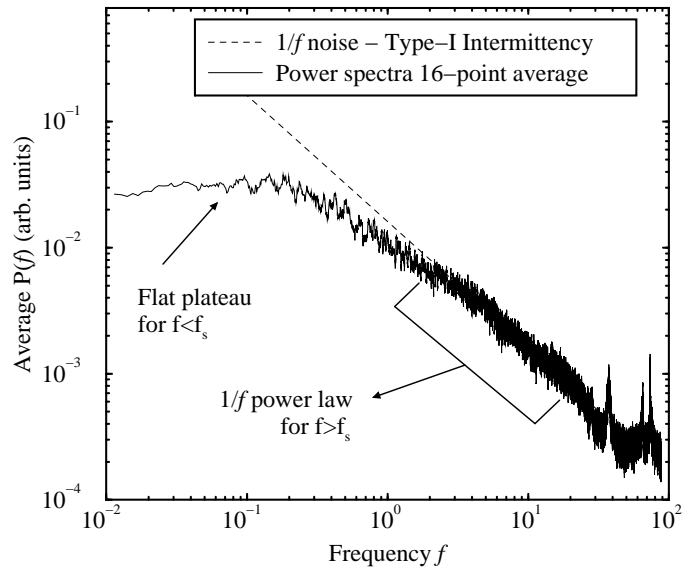}}
\caption{\label{TypeI.statistics}
Power spectra of the time series in Fig.\
\ref{TypeI.series} for Type-I intermittency.}
\end{figure}

As further evidence for this type of intermittency in the model (1), we have
plotted in Fig.\ \ref{TypeI.statistics} the power spectrum at
$C_\alpha=28.0$, obtained by averaging over 16 different initial conditions
corresponding to different initial parities. As can be seen, the power spectrum
shows both the flat plateau and the $1/f$ power law scaling.

Taken together, these indicators amount to strong evidence for the presence of
Pomeau--Manneville Type-I intermittency for this model.

\section{Conclusion}
We have obtained concrete evidence, in terms of phase space signatures,
spectra and scalings to demonstrate the presence of crisis--induced and the
Pomeau--Manneville Type-I intermittencies in axisymmetric mean--field PDE
dynamo models.  Despite the rather idealised nature of these models, this is
of potential importance since it shows the occurrence of two more types of
intermittency (in addition to in--out intermittency recently discovered
\cite{inout}) in these models which may in turn be taken as an indication that
more than one type of intermittency may occur in solar and stellar dynamos.
This suggests that any observational programme for identifying the mechanisms
underlying grand minima type variability needs to take into account the
possibility that multiple intermittency mechanisms may be operative in
different stars of the similar type, or even in the same star over different
epochs.  This would also be of importance in the interpretation of proxy data. In
this way a more appropriate hypothesis regarding such variability would be
that of {\it multiple--intermittency hypothesis}. \\

We would like to thank Axel Brandenburg for providing us with
the original code and Andrew Tworkowski for helpful discussions.
EC is supported by grant BD/5708/95 -- PRAXIS XXI, JNICT.
RT benefited from PPARC UK Grant No. L39094.

\end{document}